\documentclass[prodmode,acmtoit]{acmsmall} 
\pdfoutput=1
\usepackage{graphicx}
\usepackage{tabu}
\usepackage{booktabs}
\usepackage{makecell}
\usepackage{diagbox}
\usepackage{hhline}

\usepackage{rotating}

\usepackage{multirow}
\usepackage{color}
\usepackage{enumerate}
\usepackage{url}
\usepackage[flushleft]{threeparttable}
\usepackage{color}
\usepackage{tabularx}
\usepackage{array}
\usepackage{lineno}
\usepackage[table]{xcolor}
\usepackage{array}
\newcolumntype{?}{!{\vrule width 2pt}}

\usepackage[utf8]{inputenc}
\usepackage{longtable}
\usepackage{array}
\usepackage{eqparbox}
\usepackage{makecell, caption, booktabs}
\usepackage{helvet}

\definecolor{lightgray}{gray}{0.90}

\newcommand\Tstrut{\rule{0pt}{2.0ex}}         
\newcommand\Bstrut{\rule[-0.9ex]{0pt}{0pt}}   

\usepackage[ruled]{algorithm2e}

\SetAlFnt{\small}
\SetAlCapFnt{\small}
\SetAlCapNameFnt{\small}
\SetAlCapHSkip{0pt}
\IncMargin{-\parindent}

\begin{document}

\markboth{C. Perentis et al.}{Anonymous or not? Understanding the Factors Affecting Personal Mobile Data Disclosure}

\title{Anonymous or not? Understanding the Factors Affecting Personal Mobile Data Disclosure}
\author{CHRISTOS PERENTIS
\affil{Telecom Italia - SKIL \& Fondazione Bruno Kessler}
MICHELE VESCOVI
\affil{Telecom Italia - Semantics \& Knowledge Innovation Lab (SKIL)}
CHIARA LEONARDI 
\affil{Fondazione Bruno Kessler (FBK)}
CORRADO MOISO
\affil{Telecom Italia - Future Center}
MIRCO MUSOLESI
\affil{University College London}
FABIO PIANESI
\affil{Fondazione Bruno Kessler}
BRUNO LEPRI
\affil{Fondazione Bruno Kessler}}

\begin{abstract}
The wide adoption of mobile devices and social media platforms have dramatically increased the collection and sharing of personal information.
More and more frequently, users are called to take decisions concerning the disclosure of their personal information. In this study, we investigate the factors affecting users' choices toward the disclosure of their personal data, including not only their demographic and self-reported individual characteristics, but also their social interactions and their mobility patterns inferred from months of mobile phone data activity. We report the findings of a field-study conducted with a community of 63 subjects provided with (i) a smart-phone and (ii) a Personal Data Store (PDS) enabling them to control the disclosure of their data. We monitor the sharing behavior of our participants through the PDS, and evaluate the contribution of different factors affecting their disclosing choices of location and social interaction data. Our analysis shows that social interaction inferred by mobile phones is an important factor revealing willingness to share, regardless of the data type. In addition, we provide further insights on the individual traits relevant to the prediction of sharing behavior.
\end{abstract}

%
%

%
%



\acmformat{Christos Perentis, Michele Vescovi, Chiara Leonardi, Corrado Moiso, Mirco Musolesi, Fabio Pianesi and Bruno Lepri, YYYY. Anonymous or not? Understanding the Factors Affecting Personal Mobile Data Disclosure.}

\begin{bottomstuff}
Author's addresses: C. Perentis, Telecom Italia - Semantics \& Knowledge Innovation Lab (SKIL) \& Fondazione Bruno Kessler, Via Sommarive 18, 38123 Trento, Italy;
M. Vescovi, Telecom Italia - SKIL, Via Sommarive 18, 38123 Trento, Italy;
C. Moiso, Telecom Italia - Future Center, via Reiss Romoli 274, 10148 Torino, Italy;
M. Musolesi, Departement of Geography, University College London, Gower Street WC1E 6BT, London, United Kingdom;
C. Leonardi, F. Pianesi {and} B. Lepri, Fondazione Bruno Kessler, Via Sommarive 18, 38123 Trento, Italy. \\ 
\copyright~ACM, YYYY. This is the author's version of the work. It is posted here by permission of ACM for your personal use. Not for redistribution. The definitive version was published in PUBLICATION, \{V, N, YYYY\} \url{http://doi.acm.org/10.1145/nnnnnn.nnnnnn}
\end{bottomstuff}
\setcopyright{acmcopyright}
\issn{1533-5399/2016} 
\maketitle

\section{Introduction}
The wide adoption of mobile phones, Internet services, social media platforms, and the proliferation of wearable devices and connected objects (Internet of Things) have resulted in a massive production of personal data that characterize many aspects of daily life at extremely fine temporal and spatial granularities~\cite{lane2010survey,madan2012sensing,bettini2015}.

The availability of such a huge amount of data represents 
an invaluable resource for designing and building systems able to understand people as well as communities' needs and activities so as to provide tailored feedback and services~\cite{lathia2013smartphones}. 

At the same time, an increasing number of applications makes it easier for people 
to share their personal information (e.g., current location, activities in which they are 
involved and other contextual information) across many social networking 
applications and mobile apps~\cite{hsieh2007field,miluzzo2008sensing,tang2006putting}. 
These scenarios, however, raise unprecedented privacy challenges and concerns, with users being continuously called to take decisions concerning the disclosure of their personal information on the basis of a difficult trade-off between data protection, given the potential for user identification~\cite{de2013unique,de2015unique,rossi14:itstheway,RosMus15}, and the advantages stemming from data sharing~\cite{acquisti2015}. 

Several researchers have therefore started investigating the role of various factors in influencing the attitude towards data disclosure: e.g., interpersonal relationships~\cite{DBLP:conf/chi/ConsolvoSMLTP05,wiese2011you}; user characteristics such as gender~\cite{hoy2010gender}, age~\cite{christofides2012hey} or personality traits~\cite{quercia2012facebook,schrammel2009personality}; and the type of the shared data~\cite{knijnenburg2013dimensionality}.

Our study makes a step further in this direction. Besides considering only demographics, self-reported personality traits and privacy dispositions, our work takes into account the role played by behavioral information about social interactions and mobility patterns, extracted by the user's mobile phone. We focus in particular on the sharing of information about locations and social interactions data types.

In order to investigate all these factors, we ran a field-study with a community of 
63 subjects. They were provided with (i) a smartphone incorporating a sensing software
explicitly designed for collecting mobile phone data; and (ii) a Personal Data Store 
(PDS), a system meant to both enable subjects to raise awareness of their data and to control their disclosure with the other members of the community as well as to keep track of their actual sharing behavior.
A relevant aspect of our approach is that we observe the actual sharing behavior on real user data rather than attitudes expressed through questionnaires. 

Personal Data Stores (PDS) are systems designed to provide users with control over their personal data disclosing choices towards third-parties (e.g., on-line apps and services). More specifically, such systems enable services to access personal data and meta-data through mechanisms preserving users' privacy~\cite{mun2010personal,DBLP:conf/data/MoisoAV12,de2014openpds}. By design they are meant to create a trusted environment where several other mobile/web services, e.g., using communication, location or sensor data, interact with the user. In addition, users can actively see their data being fed to the on-line services and the potential benefit they receive from them. 

We may think about a scenario where the personal information derived from the Internet services and from the PDS can be used for the design and enhancement of privacy-preserving systems. A designer could imagine to personalize default privacy settings or to recommend sharing policies in an adaptive way by using the most informative behavioral features.

Our results show that it is possible to identify disclosing information behavioral routines by extracting features for example from call and SMS data or a PDS Internet service. In other words, we can single out key factors that can be used to understand users' privacy related behaviors. Moreover, we can highlight meaningful combinations of factors derived from mobile data, behavioral patterns of a PDS Internet service or individual characteristics that maximize the understanding of the issues related to the disclosure of personal data. Such potential could encourage the development of Internet services towards a more transparent direction.

The main contributions of this work can be summarized as follows. First, we run a field-study within a living-lab where people share continuously their real data. In this experimental setting we capture the dynamic sharing behavior of users concerning 
personal information and not just a static choice.
Second, we compute several families of features related not only to self-reported demographics, 
personality traits and privacy attitudes, but also behavioral communication 
and mobility information captured by mobile phones as well as usage patterns extracted from a PDS. 
Finally, we experimentally evaluate and highlight the effects of those factors
on the choice users make when selecting their privacy settings for two particular 
types of personal data, location and social interactions. 

\section{Related Work}\label{Sec:RelatedWork}
Previous research has considered a number of factors 
that can explain individual attitudes and preferences toward disclosing personal 
information. Demographic characteristics, such as gender and age, have been found 
to affect disclosure attitudes and behavior. Several studies have identified gender 
differences concerning privacy concerns and consequent information disclosure behaviors:
for example, women are generally more protective of their online privacy with regard 
to the amount of data disclosed on social networking platforms~\cite{hoy2010gender}. 
Similarly, in a study on Facebook usage~\citeN{fogel2009internet} found that women are less
likely than men to share personal data such as instant messenger address, home place or phone number 
on their profile page. Age also plays a role in affecting information disclosure behavior. 
For example, in a study with 288 adolescents and 285 adults on Facebook usage,~\citeN{christofides2012hey} found that adolescents disclose more information than adults. 

Prior work also emphasizes the role of personality traits - e.g., individual 
stable psychological attributes - to explain risk perception and consequent 
information disclosure behavior.~\citeN{korzaan2009demystifying}
explored the role of the Big-5 personality traits~\cite{costa2008revised} and found 
that Agreeableness, defined as being sympathetic, straightforward and selfless, 
has a significant influence on individual concerns 
for information privacy.~\citeN{junglas2008personality} 
and~\citeN{amichai2010social}, again used the Big-5 
personality traits and found that Agreeableness, Conscientiousness, and Openness to Experience affect the concern for privacy. However, other studies targeting the influence of personality traits did not find significant correlations~\cite{schrammel2009personality,massa2015}.

An interesting and extensive study is that conducted by~\citeN{quercia2012facebook} 
with 1,313 Facebook users in US. The authors investigated the role of the Big-5 
personality traits and they found weak correlations among Openness to Experience 
and, to a lesser extent, Extraversion and the disclosure attitudes on Facebook. 
In 2010,~\citeN{DBLP:conf/amcis/Lo10} suggested that Locus of Control~\cite{rotter1966generalized} 
could affect individual perception of risk in disclosing personal information, 
with internals (i.e., people who believe that their own actions merely determine their life events)
being more likely than externals (i.e., people who believe that mostly external factors determine their life events) 
to feel that they can control the risk of becoming privacy victims, hence more willing 
to disclose/share their personal information. Additional work has also showed a 
positive association between users' sociability 
captured by their personal network size and the subject's behavior with respect to information disclosure: 
subjects characterized with high sociability tend to share more information and to have less privacy concerns~\cite{young2009information}. 

Building on these findings and following the suggestions by~\citeN{jensen2005privacy}, 
our work connects demographic factors, individual traits and dispositions to the actual sharing behavior 
of people rather than attitudes expressed through questionnaires. 
Moreover, we focus our attention not only on demographic factors, individual 
traits and dispositions, 
but also on behaviors directly measured (i.e., inferred) from the data themselves
(e.g., number of calls, diversity in interactions, physical
distance traveled, etc.).

\section{Field Study}  \label{Sec:fieldstudy}
In this section we describe the methodology followed during our 15-week study.

\subsection{The Living Laboratory}
We conducted our field study within the Mobile Territorial Lab~\citeN{MTL2012}, 
a long-term living lab launched in November 2012 as a joint effort between industrial and academic research institutions~\cite{centellegher2016mobile}. 
It consists of a group of
volunteers who carry 
in their daily life
an instrumented smartphone in exchange for a monthly credit bonus of voice, SMS, and data access. 
Specifically, participants are provided with (i) an Android-based smartphone running a 
sensing software that continuously collects different types of mobile phone data 
(e.g., communication events, location, apps usage, etc.)~\cite{aharony2011social}, and (ii) a tool, called Personal 
Data Store (PDS)~\cite{de2014openpds}, which stores the participant's information and enables him/her
to exercise full control on own data management~\cite{Vescovi:2014:MDS:2638728.2638745}. By using the PDS, subjects can decide at any 
time about whether and how to disclose their data to the other participants.
One of the most important characteristics of MTL is its ecological validity, 
given that the participants' behaviors are sensed in the real world, as people 
live their everyday life, and not under artificial laboratory conditions.

All volunteers were recruited within the target group of young families with \mbox{children} using 
a snowball sampling approach where study subjects recruit future subjects from among
their acquaintances~\cite{goodman1961snowball}. Upon agreeing to the terms of participation, 
the volunteers granted researchers legal access to their behavioral data collected by their smartphones. However, volunteers retain full rights over their personal data such that they can ask to delete the 
collected information from the secure storage servers.
Moreover, participants have the choice to participate or not in a specific study.
\begin{figure*}[!t]
	\centering 
	\includegraphics[scale=0.475]{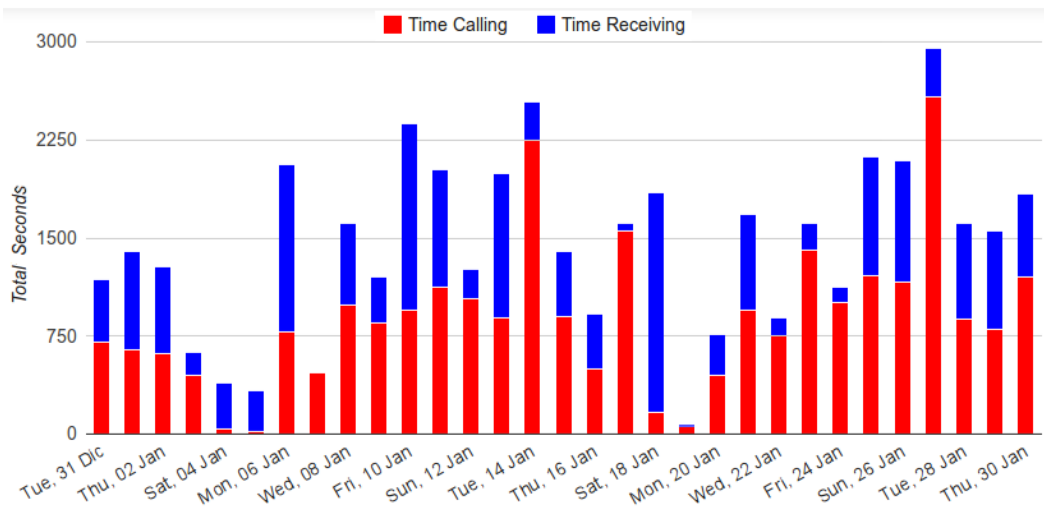}
	\caption{Individual Views: example of a PDS individual view for call interactions.}
	\label{Fig:individual_views}
\end{figure*}
In the current paper, we report a study conducted on 63 individuals (20 males and
43 females) from the MTL community. \mbox{Participants' age ranged from 28 to 46 years old} (mean = 38.67 and standard deviation = 3.34). They held a variety of occupations and education
levels, ranging from high school diplomas to PhD degrees. All were savvy Android users
who had used the smartphones provided by the living lab since 8 months before. All
participants lived in Italy and the vast majority were of Italian nationality. 
The sample is 
characterized by a medium-low social connectivity. On average
subjects declared to know 7.94 other subjects (out-degree) and resulted to be
known by 7.84 (in-degree). In the following subsections, we outline the procedure adopted for the current study
and we describe more in detail the mobile sensing platform, the PDS, and the collected
data about participants' demographic characteristics and individual traits.

\subsection{Experimental Setup}\label{experimental_setup}
The study took place for 15 weeks from July to November of 2013. 
Before the official beginning of the study, participants were asked to fill a survey 
including scales targeting: (i) Big-5 personality traits~\cite{perugini2002big}, (ii) 
Locus of Control~\cite{farma2000questionario}, (iii) Dispositional Trust~\cite{mayer1999effect}, (iv) Self-Disclosure~\cite{cozby1973self}, and
(v) privacy concerns~\cite{smith1996information}.

\begin{figure*}
	\centerline{\includegraphics[scale=0.45]{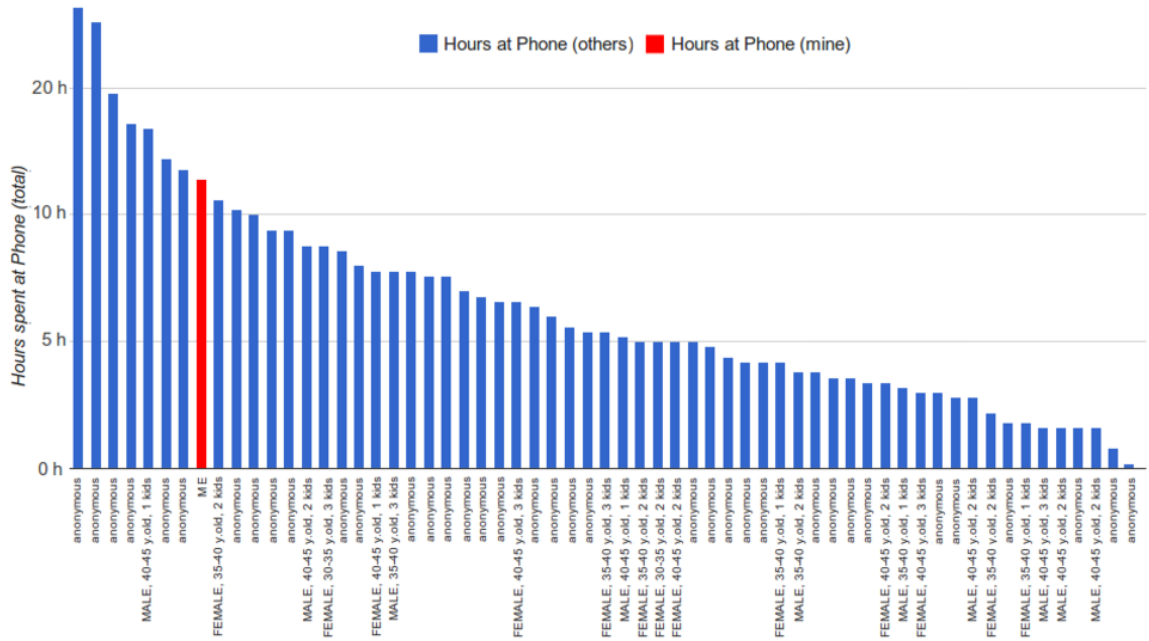}}
	\caption{Social Views: example of a PDS social view for call interactions.}
	\label{fig:social_views}
\end{figure*}

On the first day of the study, participants were asked to set their initial disclosure preferences 
on the privacy setting area provided by the PDS. From that time on, subjects were free to change their setting
at will and at any time. A week after, we started providing subjects with the social views (see Figure~\ref{fig:social_views})
built from the data disclosed in the community. Both the individual and the social views are generated by the PDS. At the end of the
study subjects were asked to set their final sharing preferences on the PDS. 

\subsection{Mobile Sensing Platform} \label{subsec: sensingplat}
The sensing software 
runs in a passive manner
and does not interfere with the normal usage of the phone. The configuration is set in
a way that battery-intensive actions (e.g., GPS and Bluetooth scans) are performed in
intervals allowing usefulness while minimizing battery consumption. The data collected
consisted of: i) call logs, ii) SMS logs, 
iii) proximity data obtained by scanning near-by phones and other Bluetooth devices and iv) location data obtained using GPS
or localized WiFi. Bluetooth and GPS scans were done every 5 minutes. Note that in this study 
we use 5-months (February to June of 2013) of collected data to compute several behavioral features.

\subsection{Personal Data Store}
The PDS is a digital space, owned and controlled through a Web interface by the user, 
acting as repository for the personal information collected during the study and offering 
every user the possibility to view, control and disclose her/his own data. Data were 
organized in ``regions'' by putting together data having a similar meaning (e.g., data 
about locations were organized in the same ``region'', independently of whether they 
were collected through GPS or a WiFi hit).

One section of the PDS was designed to provide users with visualizations of their 
(always up to date) personal data. Two types of \textit{Individual Views} were provided for each 
kind of owned data: a \textit{detailed view} (in tables or maps), where every available piece 
of raw data is represented in detail, and \textit{aggregated views} (see Figure~\ref{Fig:individual_views}) with aggregations, 
at different levels, of the personal data (e.g., charts, pies, clusters of frequent locations, 
quantity of contacts, etc.).

The PDS also features a \textit{Sharing Area}~\cite{Vescovi:2014:MDS:2638728.2638745}, a space for subjects to fix the desired disclosure level of their data, distinguished into: (i) \textit{Do Not Share}; (ii) \textit{Share Anonymously};
(iii) \textit{Share Non-Anonymously} (i.e., labeling the data with some personal demographic information).
Finally, subjects' choices are directly reflected into \textit{Social Views}, shown in Figure~\ref{fig:social_views} 
and built out of the personal data disclosed by the participants.

Social views were accessible any time by the participants on their PDS in such a way 
that any change through the sharing/disclosure settings had an immediate effect on the material displayed in them. This enabled levels of comparison of one subject's behavior 
with those of the others that depended on the subject's current sharing settings. 
In more detail: a) if for a given person, a given data type and a given time the setting 
was \textit{Do Not Share}, then the corresponding social views did not exploit the corresponding 
data and the user was prevented to access any of them; b) if the setting was 
\textit{Share Anonymously} then only their aggregated and anonymous data were made 
available in social views and they could access data only in the same format; 
c) with a \textit{Share Non-Anonymously} setting, the relevant data were presented with information 
about the subject and the latter was enabled to access all the similarly disclosed 
information by the other users.

\begin{figure*}
	\centerline{\includegraphics[scale=0.5]{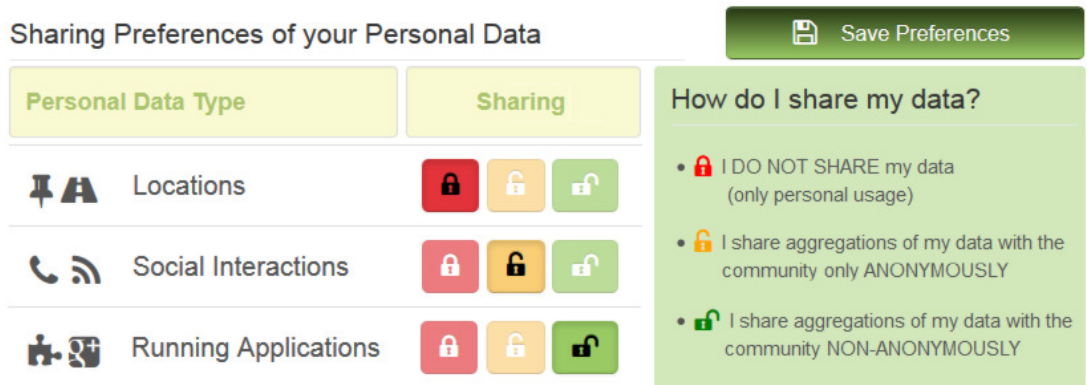}}
	\caption{PDS Sharing Area: allows users to set their PD disclosure preferences.}
	\label{fig:pds}
\end{figure*}

In summary, the level of disclosure and the social views worked 
in full synchrony to ensure that the higher the chosen disclosure level, the more detailed 
was the information made available and accessible about the others, with an increasing level 
of social comparison. To exemplify, views such as ``How much am I social?'', 
``How long I've been on the phone w.r.t. others?''. For example, Figure~\ref{fig:social_views} presents the 
latter example view for a user sharing her/his data non-anonymously; the red column 
represents the user, while on the horizontal axis the information related to the 
other users sharing ``non-anonymously'' are reported (if the user was sharing anonymously 
all the columns would be labeled as ``anonymous'').

\subsection{Demographics, Personality and Other Individual Characteristics}
We collected different types of information from our subjects including demographics,
self-reported personality traits and attitudes towards privacy. Descriptive information for the following scale scores is provided in Table~\ref{table:t2}.

\textit{Demographic Information.} As pointed out in Section 2, 
there have been several attempts to associate privacy concerns and sharing behavior with demographic information. In our case we used participants' age and gender.

\textit{Personality and Individual Traits.} In our study, Big-5 personality traits are 
measured by means of the BFMS questionnaire~\cite{perugini2002big}, 
a scale validated for Italian covering
the traditional dimensions of Extraversion, Neuroticism,
Agreeableness, Conscientiousness and Openness to 
Experience. The scale consists of 10 adjectives per personality trait, with a rating scale from 1 to 7. The Big-5 personality traits scores are obtained by summing the points of each of the 10 adjectives. We also exploited the Locus of Control (LoC)~\cite{rotter1966generalized},
a psychological construct measuring whether causal attribution for one's behavior
or beliefs is made to oneself or 
to external events or circumstances. The corresponding scale consists of a set of beliefs 
about whether the outcomes of one's actions are dependent upon what the subject does 
(\emph{internal orientation}) or upon events outside of her/his control (\emph{external orientation}). 
Locus of Control was measured by asking subjects to fill the Italian version of Craig's Locus of Control scale~\cite{farma2000questionario}. This scale is composed of 17 questions using a rating scale from 0 to 5. Each participant's Locus of Control score is computed by summing up the points of each item.

Another construct we take into account is the Dispositional Trust.~\citeN{rotter1967new} was among the first to discuss trust as a form of 
personality trait, defining interpersonal trust as a generalized expectancy that the words or promises 
of others can be relied on. In our study, we resort to~\citeN{mayer1999effect} Trust 
Propensity Scale. The Dispositional Trust scale has 8 item-questions measured in a 1 to 7 point scale. To acquire the final trust score for each subject we sum up the points of each item.

Finally, we targeted the self-disclosure attitudes of our subjects. 
Self-disclosure has been defined as any message about the self that an individual communicates to another one~\cite{cozby1973self}. 
We use Wheeless's scale, which has been utilized to measure self-disclosure in online 
communication and in interpersonal relationships~\cite{wheeless1976conceptualization}. Precisely, we measure five dimensions of self-disclosure using a 1-7 scale for each, namely: (i) amount of disclosure (7 items),
(ii) positive-negative nature of disclosure (7 items), (iii) consciously intended disclosure (4 items), (iv) honesty \& accuracy of disclosure (8 items), and (v) general depth or intimacy of disclosure (5 items). 
The final score for dimensions (iii) and (v) is the sum of points collected from the corresponding items, respectively. In contrast, for measuring dimensions (i) and (ii): 4 items sum up to $factor_a$, while 3 items sum up to $factor_b$. For computing (iv): 4 items sum up to $factor_a$ and 4 items sum up to $factor_b$. The final score for the dimensions (i), (ii) and (iv) is given each time by this construct: $(32{-}factor_a){+}factor_b$. 

\textit{Privacy Concerns.} Information about privacy concerns was collected resorting to the 
scale of Concern for Information Privacy (CFIP) developed by~\citeN{smith1996information}. 
This scale regards privacy concerns of the individual about organizational 
information privacy practices with four data-related dimensions: 
collection, unauthorized secondary use, errors and improper access to personal information.
The concerns are measured using a 1 to 7 point scale consisting of 15 question-items. The final score is computed by summing all the responses to the questions.

\textit{Social Relationships within the Community.} Each user was asked to 
indicate the people known within the community.

\section{Methodology}\label{Sec:DA}
Our goal is to understand the effect of a wide range of variables
in the disclosing decisions people make about their personal mobile data. 

To do this, we make two concrete steps. Firstly, we fit Binary Logistic Regression (BLR) models testing separately how the \emph{sharing choices} (dependent variables) are affected by the following set of independent variables' families: (i) demographic information, (ii) psychological traits and other individual dispositions (Big-5 personality traits, Locus Of Control, dispositional trust, privacy concerns, self-disclosure), (iii) social relationships within the community, (iv) dynamic behavior (communication and mobility), and (v) PDS access usage information, as visualized in Table~\ref{table:t2}. Note that testing separately per-group represents a feature selection step, since we use backward elimination. 
Secondly, taking into account exactly those features that showed a significant effect, we construct an \textit{Overall} and a combined \textit{Mobile+PDS} BLR classification model per data type in order to predict the sharing choices. The overall models represent the most effective predictors from the different families, while the combined models use only behavioral mobile data and PDS usage access features. Such data could actually be collected using a PDS service in a real-life scenario.

\begin{table}
	\tbl{Frequency Table for Initial Location \& Interactions Privacy Setting.\label{table:t1}}{
	\begin{tabular}{|l|c|c|c|c|c|} \hline
		\multicolumn{3}{|c|}{Dependent Variables} & \multicolumn{3}{|c|}{Transformed Dependent Variables}\Tstrut\Bstrut\\\hline
		Privacy Preference & Location & Interactions  & Privacy Preference  & Location & Interactions\Tstrut\Bstrut\\\hline
		Do Not Share       &        2 &            1  & \multirow{2}{*}{Share Anonymously} &    \multirow{2}{*}{21} &  \multirow{2}{*}{22} \Tstrut\Bstrut\\\cline{1-3}
		Share Anonymously  &       22 &           23 &               &   & \Tstrut\Bstrut\\\hline
		Share Non-Anonymously & 39 & 39 & Share Non-Anonymously & 39 & 39\Tstrut\Bstrut\\\hline
		Total & 63 & 63& Total & 61&  62\Tstrut\Bstrut\\\hline
	\end{tabular}}
\end{table}

\subsection{Dependent Variables: Sharing Choices} \label{dv}
To model the disclosure of personal information we construct dependent variables taking into account the final disclosing choices subjects set in the PDS, one for each different 
data type: Sharing Location and Sharing Interactions (calls \& SMS). 
As said, users were able to choose among three levels of sharing, i.e., \emph{Do Not Share}, \emph{Share Anonymously} and \emph{Share Non-Anonymously}, 
for each data type. We observe from Table~\ref{table:t1} that the \emph{Do Not Share}
choice has few occurrences concerning both the location and the social interactions data. For this reason, we discarded the data instances for the sharing choice \emph{Do Not Share} for both data types.

\subsection{Independent Variables}\label{iv}

\subsubsection{Demographics, Personality and Other Individual Characteristics}
In this paper, we take into account several characteristics of our study participants. Specifically, we focus on demographic data (age and gender), Big-5 personality traits, Locus of Control (LoC),
dispositional trust, a measure of privacy concerns and the five variables describing self-disclosure (see Table~\ref{table:t2}).

Furthermore, features regarding participants' social network were extracted using 
the self-reported information provided about the acquaintance level with 
the other people inside the community. More specifically, the following variables have been computed: (i) out-degree (i.e., the number of people that a person reports she/he knows) 
and (ii) in-degree (i.e., the number of people that reported knowing a specific person).
All variables describing individuals' characteristics are normalized scalar variables, 
except gender (female/male) being a categorical dichotomous variable.

Notice also that it was not possible to understand how the choice of a ``friend'' affects the disclosing option, because subjects are not aware of the identity of the other users' privacy setting. In the best case, if both parties share openly the data they could see each other demographic information (see Figure~\ref{fig:social_views}) but not the name. Therefore, we focused our analysis on features that characterize their social network size, i.e., in/out-degree. 

\begin{table}
	\tbl{All features included in the analysis extracted from: Self-reported information, mobile data and PDS usage.  \ \label{table:t2}}{
	\begin{tabular}{|c|c|c|c|c|c|c|} \hline\Tstrut\Tstrut
		
			\multirow{1}{*}{\textbf{Data Source}}&{\textbf{Data Category}}&{\textbf{Data Type}}& \textbf{Mean}&\textbf{SD}&\textbf{Min}&\textbf{Max}\Tstrut\Bstrut\\ \hline
	    \multirow{15}{1cm}[15pt]{Self-Reported Data through surveys} 
									&\multirow{2}{*}{Demographics}
									 & Age &38.67&3.34&28&46\\ 
	                                 && Gender &-&-&-&- \\\cline{2-7}\Tstrut 
	&	\multirow{5}{*}{Personality} & Extraversion & 39.78&10.06&16&59\\
	                               	 && Neuroticism&32.25&7.29&13&47\\
		                             && Agreeableness&49.78&6.893&35&63\\ 
	                               	 && Conscientiousness&45.94&9.84&21&61\\ 
		                             && Openness&44.52&6.75&28&56\\\cline{2-7} \Tstrut
	&	\multirow{3}{*}{Other Traits} & Locus of Control&27&10.34&9&59\\ 
									&& Trust&25.79&5.92&13&46\\ 
									&& Privacy Concerns&80.11&12.55&53&102\\\cline{2-7} \Tstrut
	&	\multirow{5}{*}{Self-Disclosure} 
									& In(Un)tentional&21.35&4.36&9&28\\ 
									&& Disclosure Amount&26.48&8.74&8&46\\ 
									&& Positive-Negative&34.19&6.34&18&46\\ 
									&& Depth-Intimacy&15.94&6.71&5&34\\
									&& Honesty-Accuracy&39.84&7.89&20&52\\\cline{2-7} \Tstrut
	&	\multirow{2}{*}{Community SN} 
									& Out-degree&7.94&4.45&2&22 \\
									&& In-degree &7.84&4.74&2&25\\	\hline\Tstrut
			
			\multirow{14}{1cm}[15pt]{Mobile Phone Data} 
						& \multirow{5}{*}{Calls} 
						& \#Total Calls&1713.71&526.84&695&2876\\ 
						&& \#Unique Call Contacts&163.4&50.59&70&339\\             
						&& Call Diversity&0.71&0.06&0.53&0.85\\
						&& avg. Calls (daily)& 12.53	&	3.74	&	5.39	&	22.29	\\         
						&& std. Calls (daily)&8.19	&	2.24	&	4.07	&	16.16	\\  \cline{2-7} \Tstrut
			& \multirow{5}{*}{SMS} 
						& \#Total SMS&1027.76&401.29&112&2036\\    
						&& \#Unique SMS Contacts&92.37&38.53&32&258\\ 				
						&& SMS Diversity&0.73&0.06&0.58&0.91\\  
						&& avg. SMS (daily)&7.90	&	2.43	&	3.20	&	14.24	\\     
						&& std. SMS (daily)&5.80	&	1.83	&	2.08	&	10.09	\\     \cline{2-7} \Tstrut
			& \multirow{4}{*}{Location}
						 & Total Distance &5604.34&2338.14&2305.94&11549\\ 
						&&std. Displacements &336.57	&	190.16	&	77.94	&	1106.25		\\     
						&& avg. Distance (daily)&40.82	&	16.11	&	16.83	&	90.94			\\     
						&&avg. std. Displ. (daily)&2.54&1.41&0.57&7.42\\                      \cline{1-7}\Tstrut
			\multirow{4}{1cm}{PDS Usage Data} 
						& \multirow{2}{*}{Location} 
						& Individual Views&1.78&1.56&0&9\\     
						&& Social Views&1.33&1.32&0&5\\						 \cline{2-7}\Tstrut
						&\multirow{2}{*}{Interactions} 
						& Individual Views&1.83&1.49&0&8\\	
						&& Social Views&1.37&1.46&0&6\\	 \cline{2-3}
	\hline

		\end{tabular}}
	\end{table}

\subsubsection{Dynamic Behavioral Data}
We computed a number of features from participants' 
mobile phone usage behavior, willing to examine if they could 
associate with personal information disclosure decisions. 
In Table~\ref{table:t2} all the behavioral features (computed over the aforementioned $5$ month period) appear combined with descriptive information. 
Firstly, we consider location and social interaction (calls \& SMS) information, 
collected passively from the mobile phone.

For both social interaction data (i.e., calls \& SMS) we compute the following five features adjusted to each data type context, as shown in Table~\ref{table:t2}. The first three concern the whole period of the study (i.e., $5$ months), while the last two ones quantify a daily behavior taking into account the days that users were actively communicating. Note that our community is really active, thus for a participant the total days of active communication is almost equal to the days of the study.
The features are the total number of calls (outgoing/incoming) and SMS (sent/received). We also consider the number (\#) of unique calls contacts and SMS contacts, and the calls' and SMS diversity.
This measure of diversity~\cite{eagle2010network} quantifies how the individuals spread their time among their contacts. More precisely, it is given by the following formula: 
\begin{equation}
\label{Eq:diversity}
{D(i) = \frac{-\sum_{j=1}^{k}p_{ij}\log{p_{ij}}}{\log{k}}},
\end{equation}
where $p_{ij}$ is the volume of communication interactions (calls or SMS) between subject $i$ and $j$ 
normalized by the total number of $i's$ calls or SMS, and 
$k$ is the distinct number of individuals contacted by calls or SMS, respectively.
High values of the diversity measure indicate that participants distribute their time more evenly among their contacts. Finally, we extract the daily average and standard deviation of the calls and SMS events, using the days when users were active.
	
To characterize participants' mobility behavior we extract metrics quantifying amount and deviation of the movement recently used by~\citeN{DBLP:conf/huc/CanzianM15}. Regarding the amount, we compute the total distance covered by the subject, i.e., the sum of the geodesic distance of the subsequent latitude and longitude coordinate pairs during the 5 months period. In addition, based on the days the user was found active we compute the daily average distance covered. Note that we exclude coordinates not matching Italy's territory for two reasons, (i) to capture everyday life behavior and (ii) to avoid outliers generated by very high distances in-between countries when traveling (e.g., by airplane). 
Those features capture the amount of mobility of a subject. Next, we measure the standard deviation of displacements, where displacement stands for the distance between one visited pair of coordinates and the subsequent one. This measure quantifies how much each location transition refrains from the total user movement. We also include the daily average for the standard deviation of displacements, quantifying a deviation of the visited locations from the average daily movement.

\subsubsection{Personal Data Store Usage}
We also investigated the role played by Personal Data Store usage by computing: (i) the total 
number of distinct days participants accessed the individual views and the social views for both location and interaction data types.
Those metrics will provide us with insights of how users
used the tool and which kind of feedback (i.e., the individual or the social one) they visited more often per data type.

\subsection{Logistic Regression Analysis and Classification}\label{str_models}
As previously mentioned, we first investigate the predictive role played by the different groups of independent variables. Then, using for each group only the factors showing a significant effect we build a combined \textit{Mobile+PDS} and an \textit{Overall} model for our two dependent variables, (i) \textit{Sharing Location} and (ii) \textit{Sharing Interactions}. Features included in all models are selected by using backward elimination. For cross-validation we use a leave-one-subject-out approach.

As evaluation metrics we report the Cox \& Snell's and the Nagelkerke pseudo $R^2$ measures to indicate the variance explained by the models. Specifically, Cox \& Snell's $R^2$ calculates the proportion of unexplained variance, which is reduced as we add more variables to the model~\cite{hardin2007generalized}. However, the maximum value of Cox \& Snell's $R^2$ can be less than $1$, making it difficult to interpret. Instead, the Nagelkerke $R^2$ varies from $0$ to $1$ (normalized Cox \& Snell's) and it is easier to interpret. The assessment of the goodness-of-fit for the models is illustrated by the Hosmer \& Lemeshow Test. It tests the hypothesis \textit{$H_0$: the model is fit}. All $p{<}0.05$ reject the null hypothesis $H_0$, meaning that the model poorly fits the data~\cite{hardin2007generalized}. Moreover, for each model we provide a classification accuracy measure (ACC) for the privacy choices \textit{Share Anonymously} and \textit{Share Non-Anonymously}. 

The strong relationships among the independent variables within each group might indicate the presence of multicollinearity. This is a common concern in regression resulting into high standard errors (${>}2$) of the $\beta$ coefficients and producing non-interpretable models with poor fit, especially for small sample sizes. For example, we notice very strong correlations ($p{>}0.95$) between the total values and the daily averaged values of the mobile features. For this reason, we test them separately, obtaining the same models. Additionally, we report significant correlations between the dichotomous categorical dependent variables and the independent by applying the Point-biserial $r_{bp}$ coefficient. 

\section{RESULTS} \label{sec:iv}

\subsection{Testing groups of predictors}  \label{sec:iv_1}

\subsubsection{Demographic Information}
Our results show that age and gender do not affect the sharing choices on location and interaction data. The $\chi^2$ independence test for both DVs using gender shows that their independence was not significantly rejected ($p{>}0.05$). Hence, we do not observe any gender difference in the sharing choice for both data types.

Concerning the \emph{Age} factor, we do not discover any association with the two dependent variables. Finally, we do not observe any significant effect in the regression tasks using the \emph{Age} and \emph{Gender} variables.

\subsubsection{Personality, Self-disclosure, and Other Traits}
Interestingly, it seems that self-disclosure affects the sharing choices, while the Big-5 personality traits and the other individual traits (i.e., dispositional trust, Locus of Control, privacy concerns) do not. Specifically, self-disclosure factors are significantly associated with the sharing choice both for location and interaction data as shown in Table~\ref{tab:t9}.
In detail, a logistic regression classifier using \emph{In(Un)tentional}, \emph{Depth-Intimacy}
and \emph{Honesty-Accuracy} provides a considerable classification gain in comparison with a baseline model using only the intercept. Indeed, it can predict the Sharing Location choice with a classification accuracy of $70.49\%$ (see Table~\ref{tab:t9}). Out of the $3$ features, only \emph{Honesty-Accuracy} presents a positive effect to the sharing choice.
 
Turning our attention to the Sharing Interactions choice, we observe that the effect is not so strong as the one observed for Sharing Location. Indeed, we classify the Sharing Interactions choice with an accuracy of $62.9\%$, equivalent to the one obtained using only the intercept (baseline model). Similarly with the location data \emph{Depth-Intimacy} presents a negative effect to the choice, but \emph{Positive-Negative} a positive one.

\begin{sidewaystable}
	\setlength\arrayrulewidth{1pt}  
	\tbl{Binary Logistic Regression Models (feature selection=Backwards Elimination) for Sharing Location and Sharing Interactions. The presented models concern \textbf{only} the groups of variables yielding a significant effect. An intercept model for each DV is included as a baseline predictor. White cells indicate that a variable has been used as an input in a BLR model. Variables per model obtaining a significant effect are fed to a cross-validated (leave-one out) logistic regression classifier. Classification Accuracy is reported for all models. \textbf{Significance: *$p{<}0.05$, **$p{<}0.01$, ***$p{<}0.001$}.}{
		\centering
		
		\begin{tabu}{|l?c|c|c|c|c|c||c|c|c|c|c|c|} \hline\Tstrut

			~&\multicolumn{6}{c||}{\textbf{Sharing Location} ($n=61$)}&	\multicolumn{6}{c|}{\textbf{Sharing Interactions} ($n=62$)}\\ \hline 
			\rowfont{\scriptsize}   
			\diagbox{Features}{Model\\Name}
			&\rotatebox[x=0cm]{90}{\scriptsize{Intercept}}
			&\rotatebox[x=0cm]{90}{\scriptsize{Self-Disclosure}}
			&\rotatebox[x=0cm]{90}{\scriptsize{Mobile}}
			&\rotatebox[x=0cm]{90}{\scriptsize{PDS}}
			&\rotatebox[x=0cm]{90}{\scriptsize{Mobile + PDS}}
			&\rotatebox[x=0cm]{90}{\scriptsize{Overall}}
			&\rotatebox[x=0cm]{90}{\scriptsize{Intercept}}
			&\rotatebox[x=0cm]{90}{\scriptsize{Self-Disclosure}}
			&\rotatebox[x=0cm]{90}{\scriptsize{Mobile}}
			&\rotatebox[x=0cm]{90}{\scriptsize{PDS}}
			&\rotatebox[x=0cm]{90}{\scriptsize{Mobile + PDS}}
			&\rotatebox[x=0cm]{90}{\scriptsize{Overall}}\\ \hline\hline
			
			\rowfont{\scriptsize}\rowcolor{lightgray}\cellcolor{white}In(Un)tenional		 &		&	\cellcolor{white}-1.364$^{*}$	&		&	&	&\cellcolor{white}	&		&\cellcolor{white}	&	&	&	&	\\
			\rowfont{\scriptsize}\rowcolor{lightgray}\cellcolor{white}Amount Disclosure		 &		&	\cellcolor{white}				&		&	&	&	&		&	\cellcolor{white}		&		&	&	&	\\ 
			\rowfont{\scriptsize}\rowcolor{lightgray}\cellcolor{white}Positive-Negative		 &		&	\cellcolor{white}				&		&	&	&	&		&\cellcolor{white}0.683$^{*}$			&		&		&&	\cellcolor{white}0.708$^{*}$	\\ 
			\rowfont{\scriptsize}\rowcolor{lightgray}\cellcolor{white}Depth-Intimacy			 &		&	\cellcolor{white}-1.328$^{**}$	&		&	&	&\cellcolor{white}-0.890$^{*}$				&		&	\cellcolor{white}-0.770$^{*}$	&		&	&	&\cellcolor{white}-0.805$^{*}$		\\  
			\rowfont{\scriptsize}\rowcolor{lightgray}\cellcolor{white}Honesty-Accuracy 		 &		&	\cellcolor{white}1.370$^{*}$	&		&	&	&\cellcolor{white}	&		&\cellcolor{white}		&	&	&	&\\ \hline

			\rowfont{\scriptsize}\rowcolor{lightgray}\cellcolor{white}\#Unique Call Contacts &		&					&\cellcolor{white}		&		&&		&		&		&\cellcolor{white}		&	&	&	\\ 
			\rowfont{\scriptsize}\rowcolor{lightgray}\cellcolor{white}Call Diversity 			&		&					&\cellcolor{white}0.769$^{*}$		&		&\cellcolor{white}$0.969^{*}$&	\cellcolor{white}$0.846^{*}$	&	&			&\cellcolor{white}0.721$^{*}$		&	&\cellcolor{white}0.874$^{*}$
			&\cellcolor{white}0.987$^{*}$			\\ 
			\rowfont{\scriptsize}\rowcolor{lightgray}\cellcolor{white}avg. Calls (daily)		&		&					&\cellcolor{white}		&		&	&	&	&	&	\cellcolor{white}	&	&	&		\\ 
			\rowfont{\scriptsize}\rowcolor{lightgray}\cellcolor{white}std. Calls (daily)		&		&					&\cellcolor{white}-0.920$^{**}$		& &	\cellcolor{white}$-1.089^{*}$		&\cellcolor{white}&			&	&	\cellcolor{white}	&	&	&	\\ 
			\rowfont{\scriptsize}\rowcolor{lightgray}\cellcolor{white}\#Unique SMS Contacts  		&		&					&\cellcolor{white}		&		&	&	&	&	&\cellcolor{white}		&	&	&	\\ 
			\rowfont{\scriptsize}\rowcolor{lightgray}\cellcolor{white}SMS Diversity  	  		&		&					&\cellcolor{white}		&		&	&	&	&			&\cellcolor{white}		&	&	&	\\  
			\rowfont{\scriptsize}\rowcolor{lightgray}\cellcolor{white}avg. SMS (daily)			&		&					&\cellcolor{white}		&	&	&		&	&	&	\cellcolor{white}	&	&	&		\\ 
			\rowfont{\scriptsize}\rowcolor{lightgray}\cellcolor{white}std. SMS (daily)			&		&					&\cellcolor{white}0.961$^{**}$	&	&\cellcolor{white}1.208$^{**}$
			&	\cellcolor{white}	&	&	&\cellcolor{white}		&	&	&	\\ 
			\rowfont{\scriptsize}\rowcolor{lightgray}\cellcolor{white}avg. Distance (daily)	&		&					&\cellcolor{white}		&		&	&	&	&			&\cellcolor{white}		&	&	&		\\
			\rowfont{\scriptsize}\rowcolor{lightgray}\cellcolor{white}avg. std. Displ. (daily) &		&				&\cellcolor{white}		&		&	&	&	&	&	\cellcolor{white}		&	&	&		\\ \hline 
			\rowfont{\scriptsize}\rowcolor{lightgray}\cellcolor{white}Social Views Location		&		&		&		&\cellcolor{white}1.142$^{**}$	&\cellcolor{white}1.412
			$^{**}$	&\cellcolor{white}1.453$^{**}$ &		&	&	&&	&			\\ 
			\rowfont{\scriptsize}\rowcolor{lightgray}\cellcolor{white}Individual Views Loc.			&		&					&		&\cellcolor{white}	&	&		&		&	&	&	&	&	\\ 
			\rowfont{\scriptsize}\rowcolor{lightgray}\cellcolor{white}Social Views Interactions			&		&					&		&	&	&		&		&	&	&\cellcolor{white}0.766$^{*}$	&\cellcolor{white}0.919$^{*}$
			&	\cellcolor{white}0.959$^{*}$	\\ 
			\rowfont{\scriptsize}\rowcolor{lightgray}\cellcolor{white}Individual Views Inter.		&		&					&		&	&	&		&		&	&	&\cellcolor{white}&	&	\\ \hline\hline
			\rowfont{\scriptsize}\textbf{Constant}														&0.572$^{*}$ 	&0.849$^{**}$&0.704$^{*}$&0.777$^{*}$&0.981$^{*}$&1.116$^{**}$ &0.528$^{*}$&0.661$^{*}$	&0.565$^{*}$&0.605$^{*}$&0.694$^{*}$   &0.892$^{*}$		\\ \hline
			\rowfont{\scriptsize}\textbf{Cox \& Snell} $R^2$ 			&				&0.265	 &0.181		&	0.159	&	0.339	&0.321&	&0.160&	0.097&	0.095	&0.195&	0.304		\Tstrut	\\ \hline
			\rowfont{\scriptsize} \textbf{Nagelkerke} $R^2$ 			&				&0.363	 &0.249		&0.219		&	0.464	&	0.441&	&0.219&	0.132&	0.129&0.266
			&0.415			\Tstrut	\\ \hline
			\rowfont{\scriptsize} \textbf{Hosmer \& Lemeshow} 					&				&0.136	 &0.244		&0.42			&	0.482&	0.487&	&0.188&	0.660&	0.472&0.821	&	0.421		\Tstrut\\\hline
			\rowfont{\scriptsize} \textbf{ACC} (\%) (CV) 			&63.90	&\textbf{70.49}   &	\textbf{67.21}	&62.29&	\textbf{73.77}&\textbf{78.68}&	62.90&	62.90&	59.68&	64.52&\textbf{72.58}&	\textbf{75.80}	\Tstrut\\\hline
		\end{tabu}}
		\label{tab:t9}
	\end{sidewaystable}
	
Interestingly, the \emph{Depth-Intimacy} factor associates negatively with the Sharing Location ($r_{bp}{=}-0.373^{*}$) and Sharing Interactions DVs ($r_{bp}{=}-0.297^{*}$) revealing a consistent tension. 

\subsubsection{Community Social Network (CSN)}
\emph{In-degree} and \emph{out-degree} variables do not capture significant effects in our regression tasks.

\subsubsection{Dynamic Behavior}
Our results show that communication factors have an impact on the sharing choices of our study participants, while the mobility ones do not yield a significant effect. Specifically, the \emph{Call Diversity} factor associates significantly 
with the Sharing Location choice ($r_{bp}{=}0.295^{*}$) and with the Sharing Interactions choice ($r_{bp}{=}0.312^{*}$). Other factors significantly associating with Sharing Location choice were the standard deviation of the daily number of calls and the standard deviation of the daily number of SMS.

A logistic regression classifier using the three aforementioned factors classifies correctly $67.2\%$ of the sharing choices. In detail, a unit increase in \emph{Call Diversity} ($e^\beta{=}2.157$) or in the standard deviation of the daily number of SMS (i.e., $e^\beta{=}2.002$ for \emph{std. SMS}) indicates an increase in the odds of sharing more openly by $115.7\%$ and $100.2\%$, respectively. In contrast, the standard deviation of the daily calling behavior (i.e., \emph{std. Calls}) has an opposite effect implying a decrease in the odds of $60.1\%$.

For the Sharing Interactions choice, the regression model keeps only \emph{Call Diversity} after a backward elimination step (see Table~\ref{tab:t9}). The remaining factors extracted from mobile phone data do not contribute to the model. More specifically, a unit increase in \emph{Call Diversity} ($e^\beta{=}2.057$) increases the probability that somebody chooses to share non-anonymously by $105.7\%$. The model classifies correctly the $62.9\%$ of the cases, equally to the baseline intercept model.

\subsubsection{PDS Usage}
The last group of factors contains features describing the usage of the PDS. Interestingly, the usage of the PDS significantly associates with the sharing choices. We find a significant positive correlation between the \textit{Social Views Location} variable and the Sharing Location choice ($r_{bp}{=}0.374^{**}$). Thus, it seems that more visits someone pays to the social views section of the PDS, the more openly s/he is about to share the location data. However, the logistic regression model using \textit{Social Views Location} as a unique factor predicts correctly only the $62.29\%$ of the sharing choices.

For the interaction data, the behavior is similar to the one observed for the location data, i.e., only \textit{Social Views Interactions} shows a positive correlation with Sharing Interaction DV ($r_{bp}{=}0.374^{**}$). The regression model performs slightly better ($64.52\%$ of accuracy) than the baseline. 

\subsection{Testing combinations: Mobile+PDS and Overall Models}\label{sec:results}
In this section, we report the results obtained using (i) the \textit{Mobile+PDS model} and (ii) the \textit{Overall model}. The former represents the exploitation of behavioral data (e.g., communication and mobility behaviors) collected by many online services, while the latter the usage of all relevant information to predict the sharing choice. 

\subsubsection{Mobile+PDS}
The \textit{Mobile+PDS model} classifies correctly the $73.77\%$ of the Sharing Location choices, and thus the combination of mobile phone data and PDS usage outperforms a model using only a single data source (e.g., only mobile phone data). 
As shown in Table~\ref{tab:t9}, a unit increase in \emph{Call Diversity} ($e^\beta{=}2.63$), in the daily \emph{std. SMS} ($e^\beta{=}3.6$) and in the \emph{Social Views Location} ($e^\beta{=}4.1$) variables increases the probability that a user shares more openly the data by $163\%$, $260\%$ and $310\%$, respectively. Practically, this means that if people distribute more evenly their call interactions among their contacts, increase the deviation in their daily SMS communication activities and increase the number of visits (distinct days) to the social views section of the PDS, they are more prone to select the \emph{Share Non-Anonymously} choice.
Instead, an increase in the standard deviation of the daily number of calls (\emph{std. Calls}) results in a $66\%$ decrease in the chance to share more openly ($e^\beta{=}0.34$). 

Turning our attention to Sharing Interactions, the \textit{Mobile+PDS} model significantly outperforms both the outcomes of the mobile phone data model and of the PDS model (\textit{Mobile+PDS} model obtain a classification accuracy of $72.58\%$).
Moreover, our results show that if we have a unit of increase in \emph{Call Diversity} ($e^\beta{=}2.4$) and in the number of visits to the PDS social views section ($e^\beta{=}2.5$), the probability that a user shares more openly her/his interaction data increases by $140\%$ and $150\%$, respectively.

\subsubsection{Overall}
The \emph{Overall} model captures the $44.1\%$ of the normalized variance and obtains an accuracy value of $78.68\%$ for Sharing Location choice. As shown in Table~\ref{tab:t9}, the factors having a significant effect are \emph{Depth-Intimacy} ($e^\beta{=}0.4$), \emph{Call Diversity} ($e^\beta{=}2.3$) and the \emph{Social Views Location} ($e^\beta{=}4.2$). More in detail, a unit of increase in the depth somebody shares information (i.e., a person scoring higher score in \emph{Depth-Intimacy} tends to  share personal information more in depth within its social circle), decreases $60\%$ the probability of sharing more openly. On the opposite, if a study participant increases the diversity of its calls and its number of visits (distinct days) to the social views functionality of the PDS, s/he increases $2.3$ and $4.2$ times, respectively, the probability of sharing openly the location data. 
 
For the Sharing Interactions choice, the \textit{Overall} model captures $44.15\%$ of the normalized variance and classifies correctly the $75.80\%$ of the choices. In particular, \emph{Call Diversity} ($e^\beta{=}2.03$) and \emph{Social Views Interactions} variables ($e^\beta{=}2.61$) show a positive effect, while the \emph{Depth-Intimacy} ($e^\beta{=}0.45$) a negative one. Moreover, the \emph{Positive-Negative} factor of the self-disclosure scale shows a significant positive effect ($e^\beta{=}2.03$). High values of this scale indicate that people disclose more positive information about themselves within their social circle than negative one. So, a unit increase in this feature increases approximately $103\%$ the probability of sharing the interaction data more openly. To exemplify, people who release more positive information to others, are more 
likely to share more openly their social interaction data. 

\section{Discussion and Conclusions}
In this section we discuss the theoretical and the practical implications of our work, and the questions that remain open.

\subsection{Theoretical Implications}
Our findings suggest that users' communication  interactions, as inferred through mobile phone data, may provide very useful information to describe users' privacy choices in disclosing personal data, thus confirming the relevance of smartphone data for understanding human behavior. 

We have shown that the diversity of calls is an important factor for predicting the sharing choice for the different data types. Interestingly, our results are in line with those recently 
obtained by~\citeN{staiano2014money} in a study designed to investigate the monetary value people assign to their personal data collected by mobile phones. 
Indeed, they identified differences in the bidding 
behaviors of the individuals which are not correlated with socio-demographic or personality traits (with the exception of Agreeableness), 
but they are correlated with behavioral differences inferred from mobile phone activity.

Regarding the individual characteristics, Depth-Intimacy
and Positive-Negative disclosure features
constitute useful information to detect users' willingness
to share personal data. Specifically, we observed (for both data types) a negative effect of Depth-Intimacy on sharing data more openly. This means that if subjects disclose more in depth information within their social circle (i.e., higher feature values), it is less likely to share communication and location information in a non-anonymous way (i.e., high disclosure) in our setting. 
Conversely, we found that people sharing more positive information with their contacts (i.e., higher Positive-Negative feature values), tend to disclose more information about their communication activity (i.e., the social interactions) in a non-anonymous way (i.e., higher disclosure). 

On the other hand we do not observe any Big-5 
personality trait affecting the disclosure of personal information, 
suggesting that such decisions do not necessarily pass through our personality mechanism. 
Previous studies on the influence individual traits 
(usually Big-5 and Locus of Control) have
on privacy attitudes and privacy-related behaviors provide contrasting 
evidence: some of them found weak correlations~\cite{DBLP:conf/amcis/Lo10,quercia2012facebook}, 
while others~\cite{schrammel2009personality} found no significant correlations. 
Hence, our results support more the latter study, but also strengthen those findings since we have obtained them within an experimentation involving a community of people sharing their data during their daily lives. However, we should argue that additional investigation is needed to clarify the role of individuals' characteristics on personal data disclosure decisions. 

Regarding the relationship between the PDS tool usage and the disclosing decisions, we found that a higher number of visits to the \textit{Social Views} functionality corresponds to a higher probability for such users to share both their location and social interactions information in a non-anonymous way. This observed effect is very strong in our case and could be interpreted as people having a higher propensity to share their information, if they receive useful feedback. An example can be users sharing their location information in a low-risk privacy-preserving way, if it would be useful for environmental studies in their territory (e.g., within the MTL project we launched an air quality crowd-sensing campaign using location sharing). Nevertheless we must keep in mind that our experimental community environment provides subjects with a sort of confidence (i.e., the perception of a more protected environment), which may lead to increased sharing. In addition, it has been previously shown that providing users with control over their data could raise their awareness, but it also might increase the perception of control (thus trust), resulting in more sharing~\cite{brandimarte2013misplaced}.

\subsection{Practical Implications} 
The main area in which our findings could be practically applied in the short term is privacy protection. Our results suggest that, by simply using (i) communication interactions (e.g., call diversity), (ii) a limited amount of user input (e.g., some information about self-disclosure dispositions), or (iii) usage patterns (e.g., PDS usage logs); a privacy-recommendation tool could offer a preliminary way of personalizing default privacy settings, which users could then change. 

More precisely, this tool may work as a layer of a Personal Data Store that activates some personalized privacy settings once third-party apps or online services perform data requests. In this way, the user is not called to continuously take decisions concerning the disclosure of her/his personal information. In addition, the exploitation of usage patterns (i.e., logs from PDS tools) could enhance a privacy setting recommendation by employing an auxiliary information source~\cite{DBLP:conf/pkdd/SymeonidisP14}.
Concerning other available sources in a recommendation task, communication diversity and self-disclosure can be computed by using digital traces or by asking simple questions, respectively.
 
Our findings may also be useful to devise a tool able to inform users about the extent to which they are exposing information that is generally considered to be sensitive by their social contacts or by people with similar personalities (e.g., similar self-disclosure characteristics) and behaviors (e.g., similar communication interactions). In this way, the tool may raise the awareness of the user about her/his sharing activity. Under this perspective, the frequency of the visits to a potential personal data intensive service (e.g., a social networking service) could be an index used to trigger notifications to the user about critical disclosing actions.

It is worth noticing that nowadays different tools and services for the user-centric personal data management are emerging~\cite{mun2010personal,DBLP:conf/data/MoisoAV12,de2014openpds,Vescovi:2014:MDS:2638728.2638745}. This is a consequence of the increased public attention the privacy concerns and issues receive. These concerns arise mainly from the increasing ubiquitous collection of personal data and from the unprecedented daily privacy challenges users face making constantly decisions between service usage and data protection~\cite{DBLP:conf/www/PerentisVL15}. Such concerns have been recently taken into account by policy-makers, as shown by the new EU directive for privacy~\cite{eu2016directive}. Starting from 2018, the reform of the EU General Data Protection Rules (GDPR) has the goal to introduce, within the whole EU, new principles such as (i) privacy by design and by default~\cite{langheinrich2001privacy}, (ii) the right of owning a copy of the personal data, and (iii) the right to be forgotten~\cite{rosen2012right}. In this study, our findings on the factors affecting personal data disclosure could provide useful insights about the development of solutions for managing personal data.

\subsection{Limitations} 
The study was conducted in a real-setting providing us the opportunity to  understand human behavior by a combination of self-reported data, mobile phone data and a real system allowing users to make decisions for the disclosing of personal information. 
However, it is important to report some limitations of the study. 
Firstly, the collection of self-reported data is always prone to bias.
Secondly, despite the high level of technical support for gathering mobile data, battery limitations or other issues might interfere with data collection.
Sample size is always a concern in such research tasks, 
but the experimental evaluation presented in this work takes into consideration
potential issues like multicollinearity, leading to non-interpretable models. The community environment and the usage of the
tool should have offered a sort of trust leading users to share their data. Finally, the sharing option may be further divided into different levels such as sharing with a specific member of our community, with people with similar behaviors or dispositions, with friends, etc. Future planned modifications on the PDS platform will add these levels to the sharing option.\\

Despite the aforementioned limitations, 
our findings open interesting directions for designing systems 
able to support users in their decisions about information 
disclosure and, more in general, 
to improve the experience of sharing personal information~\cite{knijnenburg2013helping}. 
As discussed above, a better understanding of factors affecting sharing decisions may support the design of 
adaptive systems able to suggest preselected configurations of privacy options to users, thus relieving them from the task of defining them~\cite{wiese2011you}. 

\bibliographystyle{ACM-Reference-Format-Journals}
\bibliography{bilbiography}



\medskip
\end{document}